\documentclass[twocolumn, reprint]{revtex4-2}
\usepackage{natbib}
\usepackage{graphicx}
\usepackage{amsmath}
\usepackage{gensymb}
\usepackage{color}
\usepackage{url}

\begin{document}
\title{Photoemission chronoscopy of the Iodoalkanes}
\author{Christian A. Schröder${}^{1, \dagger}$, Maximilian Pollanka${}^{1}$, Pascal Freisinger${}^{1}$, Matthias Ostner${}^{1}$, Maximilian Forster${}^{1}$, Sven-Joachim Paul${}^{1}$ and Reinhard Kienberger${}^{1}$\\
\small{${}^{1}$Chair for Laser- and X-ray physics E11, Department of Physics, TUM School of Natural Sciences, Technische Universität München, James-Franck-Str. 1,85748 Garching\\${}^{\dagger}$Contact author: \url{christian.schroeder@tum.de}}}

\begin{abstract}
Time delays in photoemission are on the order of attoseconds and have been experimentally determined in atoms, molecules and solids. Their magnitude and energy dependence are expected to yield fundamental insights into the properties of the systems in which they're measured. In a recent study Biswas \textsl{et al.} (Biswas, S., Förg, B., Ortmann, L. et al. Probing molecular environment through photoemission delays. Nat. Phys. 16, 778–783 (2020)) determined the absolute photoemission time of the I$4d$ level in iodoethane via attosecond streaking spectroscopy, finding the presence of a functional group to increase the photoemission time delay, suggesting a correlation between the size of the functional group and time delay based on a semi-classical calculation. Here we experimentally study the dependence of the I$4d$ photoemission time on the functional group in the iodoalkanes from iodomethane up to 2-iodobutane at three photon energies across the giant resonance in the I$4d\to\varepsilon f$ photoemission channel, finding that the presence alone of a functional group does not necessarily increase the photoemission delay, and that overall no clear correlation between its size and the photoemission time delay can be established.
\end{abstract}

\maketitle

\section{Introduction}

Photoemission chronoscopy, i.e. the measurement of the absolute time delay between the interaction of one (typically) extreme ultraviolet (XUV) photon with an atom, molecule or condensed matter sample and the appearance of the ejected photoelectron in vacuum, has recently been demonstrated as a viable experimental technique in a landmark experiment \cite{ossiander2018absolute}. Experimentally this scheme relies on the attosecond streak camera technique \cite{itatani2002attosecond} and measures the photoemission time of the system under study relative to a chronoscope species in which the photoemission time delay is known, e.g. from \textsl{ab-initio} calculation. \citet{ossiander2018absolute} established this scheme with the ultimate goal to determine the photoemission time of the $4f$ and valence states of a W(110) surface at an excitation energy of $105\,\mathrm{eV}$, by constructing a chain of references involving the I$4d$ photoemission in small iodoalkanes as an important chainlink. Iodine and its compounds are especially suited for this purpose due to the giant dipole resonance (GDR) in the I$4d\to\varepsilon f$ (a transition from the I$4d$ into a continuum $f$ wave with energy $\varepsilon$) channel in this energy range, which greatly enhances the photoemission cross section (cf. \cite{olney1998quantitative}), permitting their use even at very low concentrations or when XUV photons are scarce. Furthermore, the $d\to f$ photoemission channel strongly dominates the photoelectron signal when compared with the $d\to p$ channel, vastly simplifying the interpretation of the results. Following the experiment by \citet{ossiander2018absolute}, \citet{biswas2020probing} aimed at uncovering molecular effects in the observed time delay, finding that the I$4d$ photoemission time in iodoethane may be raised significantly by the presence of a functional group when compared to the atomic case, and they present a calculation that predicts that this increase should continue as the size of the functional group grows. Here we expand upon this experiment and explicitly test whether this prediction can be generalized by measuring the I$4d$ photoemission time in primary and secondary iodoalkanes from iodomethane up to 2-iodobutane at photon energies of $90\,\mathrm{eV}$, $105\,\mathrm{eV}$ and  $118\,\mathrm{eV}$. We compare our experimental results to those of \citet{biswas2020probing}, a published calculation for atomic iodine as well as a scattering calculation on the Hartree-Fock level of theory.

\section{Experimental}
\begin{figure}[ht!]
	\includegraphics{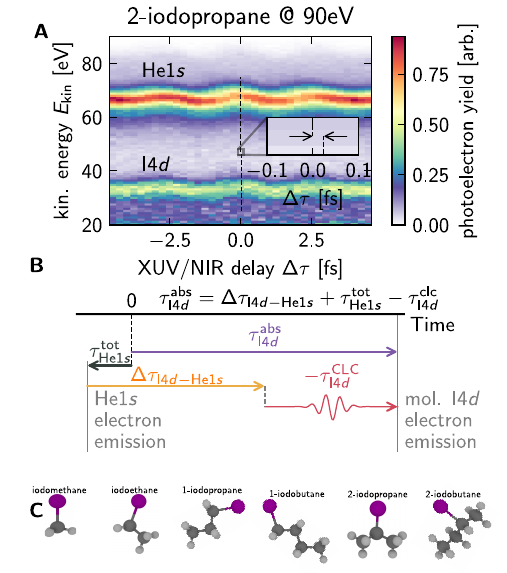}
	\caption{Experimental determination of the absolute time delay of the I$4d$ photoemission in the iodoalkanes. \textbf{A} Exemplary spectrogram recorded in a mixture of 2-iodopropane and Helium at $90\,\mathrm{eV}$ photon energy. The relative photoemission time $\Delta\tau_{\mathrm{I}4d-\mathrm{He}1s}$ is the lateral shift of the two streaking features with respect to each other and is extracted from the spectrogram via a multivariate fitting procedure. \textbf{B} With $\Delta\tau_{\mathrm{I}4d-\mathrm{He}1s}$ experimentally determined the absolute photoemission time of the I$4d$ level $\tau_{\mathrm{I}4d}^\mathrm{abs}$ can be determined via the illustrated timing scheme. The photoemission time of the He$1s$ photoemission is known \cite{ossiander2017attosecond} and can be calculated with great precision, and the measurement-induced $\tau_{\mathrm{I}4d}^\mathrm{clc}$ can be calculated analytically \cite{nagele2011time}. \textbf{C} Ball-and-stick models of the iodoalkanes under study. Iodine atoms are rendered in violet, Carbon atoms in dark gray and Hydrogen atoms in light gray.}
	\label{fig:experimental}
\end{figure}
We apply attosecond streaking spectroscopy (c.f. \cite{kienberger2004atomic, schultze2010delay, yakovlev2010attosecond}) to measure the I$4d$ photoemission time delay in iodomethane, -ethane 1- and 2-iodopropane as well as 1- and 2-iodobutane in mixture with Helium, using the known He$1s$ photoemission time delay as our reference \cite{ossiander2017attosecond, pazourek2012attosecond}.
Briefly, high-harmonics of a waveform-controlled near-infrared / visible (NIR/VIS) few-femtosecond laser pulse centered around $\lambda_\mathrm{c} = 790\,\mathrm{nm}$  are generated in a short gas cell ($500\,\mu\mathrm{m}$) backed with $\sim 100\,\mathrm{mbar}$ of Neon.
The harmonic radiation is spatially separated from the NIR/VIS light via a thin Zr foil ($0.15\,\mu\mathrm{m}$ thickness) placed at the center of the two coaxial beams after multiple meters of propagation. The spatially separated beams are then reflected off an interferometric double-mirror assembly consisting of an annular silver focusing mirror which reflects the NIR/VIS light and a focusing XUV multilayer reflector acting as a bandpass filter, which is inset into the annular silver mirror.
Via this bandpass filter XUV attosecond pulses centered at $90\,\mathrm{eV}$, $105\,\mathrm{eV}$ or $118\,\mathrm{eV}$ and $5-6\,\mathrm{eV}$ bandwidth ($\sim 300\,\mathrm{as}$ duration) are isolated from the spectrally unmodulated cut-off portion of the high-harmonic spectrum, where the photon energy is set by the design of the XUV reflector in use.
The interferometric time delay between XUV and NIR pulse is controlled by axial translation of the XUV reflector w.r.t. the outer NIR mirror via a piezo actuator.
The double mirror delay assembly focuses both beams near the exit of an effusive nozzle through which the mixture of Helium and the molecule is injected into the vacuum experimental apparatus.
In the interaction region the attosecond XUV pulse initiates photoemission, ejecting a photoelectron wave packet from a molecule or Helium atom, which is then modulated in its kinetic energy by the action of the temporally offset NIR/VIS laser field.
This modulation maps the temporal properties of the ejected photoelectron wave packet onto spectral information (cf. \cite{kienberger2004atomic, yakovlev2010attosecond}).
Photoelectron spectra are recorded with a time-of-flight electron spectrometer with an acceptance cone half-angle of $\sim 20\degree$, and a sequence of spectra recorded as a function of XUV/NIR delay in this configuration constitutes a photoelectron spectrogram.
Figure \ref{fig:experimental}A shows an exemplary spectrogram, recorded at a photon energy of $90\,\mathrm{eV}$ on a gas mixture of Helium and 2-iodopropane.
In a spectrogram the photoemission time $\Delta\tau_{\mathrm{I}4d - \mathrm{He}1s}$ of the I$4d$ level relative to that of the He$1s$ photoemission is encoded as small a lateral shift of the I$4d$ streaking feature with respect to He$1s$ feature (cf. \cite{nagele2011time, pazourek2015attosecond}, inset in fig. \ref{fig:experimental}A). No less than 30 spectrograms per photon energy and molecule are taken. For extracting $\tau_{\mathrm{I}4d - \mathrm{He}1s}$ we apply a multivariate fitting procedure, which models the streaking process based on the strong-field-approximation of the time-dependent Schrödinger equation (cf. \cite{schultze2010delay, ossiander2017attosecond, riemensberger2019attosecond, brunner2022deep, neppl2012attosecond} and supplementary information) to each recorded spectrogram to determine this time shift at the central photon energy of the XUV pulse. The entire delay extraction procedure and statistical evaluation is detailed in the supplementary information.

\begin{table}
    \caption{Absolute photoemission time delays of the I$4d\to\varepsilon f$ transition in the different molecules as determined by the experiment as well as the $\tau^\mathrm{clc}_{\mathrm{I}4d}$ contribution and the absolute time delay $\tau^\mathrm{tot}_{\mathrm{He}1s}$ of the He$1s$ photoemission (see supplementary materials). The atomic I$4d$ EWS time delay, extracted from  the publication of \citet{pi2018attosecond} is given for reference.}
    \label{tab:delays}
    \begin{tabular}{llll}
    \hline
                    & $90\,\mathrm{eV}$             & $105\,\mathrm{eV}$             & $118\,\mathrm{eV}$             \\
    \hline
    I-methane     & $\left(76.9^{+5.3}_{-11.6}\right)\,\mathrm{as}$ & $(40.9 \pm 4.1)\,\mathrm{as}$ & $(40.1 \pm 4.0)\,\mathrm{as}$ \\
    I-ethane     & $\left(62.8^{+6.2}_{-12.5}\right)\,\mathrm{as}$ & $(39.6 \pm 7.1)\,\mathrm{as}$ & $(25.1 \pm 3.3)\,\mathrm{as}$ \\
	1-I-propane & $\left(36.6^{+4.1}_{-10.5}\right)\,\mathrm{as}$ & $(23.1 \pm 7.2)\,\mathrm{as}$  & $(10.3 \pm 5.6)\,\mathrm{as}$  \\
	2-I-propane   & $\left(78.5^{+6.7}_{-13.0}\right)\,\mathrm{as}$ & $(63.9 \pm 7.1)\,\mathrm{as}$ & $(20.0 \pm 5.7)\,\mathrm{as}$ \\
	1-I-butane  & $\left(33.0^{+6.8}_{-13.1}\right)\,\mathrm{as}$ & $(31.7 \pm 6.4)\,\mathrm{as}$  & $(35.4 \pm 7.0)\,\mathrm{as}$  \\
	2-I-butane  & $\left(28.8^{+4.4}_{-10.7}\right)\,\mathrm{as}$ & $(35.3 \pm 6.2)\,\mathrm{as}$  & $(32.0 \pm 6.0)\,\mathrm{as}$  \\
    \hline
    $\tau^\mathrm{tot}_{\mathrm{He}1s}$              & $-6.5\,\mathrm{as}$           & $-5.1\,\mathrm{as}$            & $-4.3\,\mathrm{as}$            \\
    $\tau^\mathrm{clc}_{\mathrm{I}4d}$               & $-19.2\,\mathrm{as}$          & $-12.2\,\mathrm{as}$           & $-9.1\,\mathrm{as}$            \\
    $\tau^\mathrm{at.}_{\mathrm{I}4d}$ \cite{pi2018attosecond} & $32.0\,\mathrm{as}$           & $29.2\,\mathrm{as}$            & $25.8\,\mathrm{as}$            \\
    \hline
    \end{tabular}
\end{table}

\section{Results \& Discussion}
\label{sec:results_and_discussion}
Following references \cite{ossiander2018absolute} and \cite{biswas2020probing}, we take the total delay 
\begin{equation}
	\tau_{tot} = \tau_\mathrm{abs} + \tau_\mathrm{clc}
	\label{eq:total_delay}
\end{equation}
of direct photoemission as measured by attosecond streaking spectroscopy as a sum of an absolute time delay $\tau_\mathrm{abs}$ and the Coulomb-Laser-Coupling (CLC) contribution $\tau_\mathrm{clc}$, which is introduced due to the interaction of the ejected photoelectron wave packet with the streaking laser field \cite{nagele2011time, pazourek2015attosecond} in the long-ranged coulombic part of the photoion's potential. The CLC contribution is independent of the system under study and only depends on the kinetic energy of the photoelectron and the streaking pulse's wavelength. In the above expression, we identify the absolute delay $\tau_\mathrm{abs}$ as the Eisenbud-Wigner-Smith (EWS) time delay due to the scattering of the outgoing photoelectron wave packet off the potential of the photoion (c.f. \cite{ossiander2018absolute, biswas2020probing}). 

The relative I$4d-$He$1s$ time delay as measured in the experiment is then
\begin{equation}
 \Delta\tau_{\mathrm{I}4d - \mathrm{He}1s} = \left(\tau^\mathrm{abs}_{\mathrm I 4d} + \tau^\mathrm{clc}_{\mathrm I 4d}\right) - \tau^\mathrm{tot}_{\mathrm{He}1s}.
 \label{eq:rel_delay}
\end{equation}
For the direct Helium $1s$ photoemission, the total photoemission time $\tau^\mathrm{tot}_{\mathrm{He}1s}$ is known with great precision from \textsl{ab-initio} calculations in excellent agreement with experimental observation \cite{pazourek2012attosecond, ossiander2017attosecond}, and therefore eq. \ref{eq:rel_delay} can be used to calculate absolute photoemission times $\tau_{\mathrm I 4d}^\mathrm{abs}$ from the experimentally determined relative time delays. This procedure is illustrated in panel B of Fig. \ref{fig:experimental}. 

\begin{figure}[ht!!]
	\includegraphics{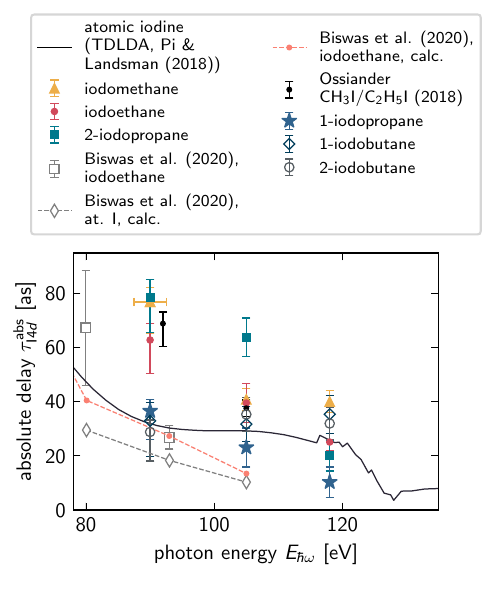}
	\caption{Absolute photoemission time delays of the I$4d$ photoelectrons in the iodoalkanes in comparison with the results of other experiments \cite{ossiander2018absolute, DissMarcus, biswas2020probing}. The black curve represents the atomic calculation by \citet{pi2018attosecond}, gray diamonds are the semi-classical calculation for atomic iodine, and salmon points are the molecular calculation from \cite{biswas2020probing}. Horizontal error bars for our data are given by the XUV reflector's bandwidth and only plotted once for clarity.}
	\label{img:delays}
\end{figure}

\subsection{Absolute photoemission times}
Turning first to iodomethane, and -ethane, we compare our experimentally determined $\tau^\mathrm{abs}_{\mathrm I 4d}$ with those determined for iodomethane and -ethane by \citet{ossiander2018absolute} and in \cite{DissMarcus} and those by \citet{biswas2020probing}. We find excellent agreement of our data for iodomethane and -ethane at $105\,\mathrm{eV}$ with the results of \citet{ossiander2018absolute} and at $90\,\mathrm{eV}$ with those from \cite{DissMarcus} at $92.5\,\mathrm{eV}$ (see Fig. \ref{img:delays}). Our data at $90\,\mathrm{eV}$, however, does not reproduce the experimental result of \citet{biswas2020probing} at $93\,\mathrm{eV}$, who use the $2p$ photoemission of Neon as their chronoscopy reference for which they calculate the absolute delay. It is possible that the difference in reference species and/or data analysis method is the reason for the disagreement betwen their results and ours.

Overall we find the photoemission times to decrease with energy in all molecules except for 1-iodobutane and 2-iodobutane where it is almost constant within the uncertainties. The decrease with energy is expected as it is a general property of the EWS delay time \cite{wigner1955lower}. Compared to a calculation for atomic iodine by \citet{pi2018attosecond} we find the photoemission time delays in iodomethane, -ethane and 2-iodopropane to be significantly elevated at at $90\,\mathrm{eV}$ and $105\,\mathrm{eV}$. The delay in 1-iodopropane exceeds the atomic calculation at $90\,\mathrm{eV}$ but is lower at the higher energies. In 1- and 2-iodobutane the atomic prediction is only exceeded significantly at $118\,\mathrm{eV}$. At no photon energy does the calculation for iodoethane of \citet{biswas2020probing} (salmon points in Fig. \ref{img:delays}) agree with our experimental data or that of \citet{ossiander2018absolute}. Of note is also the comparison between the atomic calculation of \cite{pi2018attosecond} based on the time-dependent local density approximation (TDLDA) and the semi-classical calculation for atomic iodine of \cite{biswas2020probing}: the semi-classical calculation is tens of attoseconds below the TDLDA prediction (which can be expected to be quite accurate, as it produces a photoemission cross-section in accord with spectroscopic data, e.g. \cite{nahon1991experimental, osullivan1996trends, olney1998quantitative}) for all photon energies, raising the question of whether it is actually capable of accurately describing the I$4d$ photoemission in this energy range, in particular the giant resonance.

\subsection{Variation of the photoemission time with the size of the functional group}

\begin{figure}[hb!]
	\includegraphics{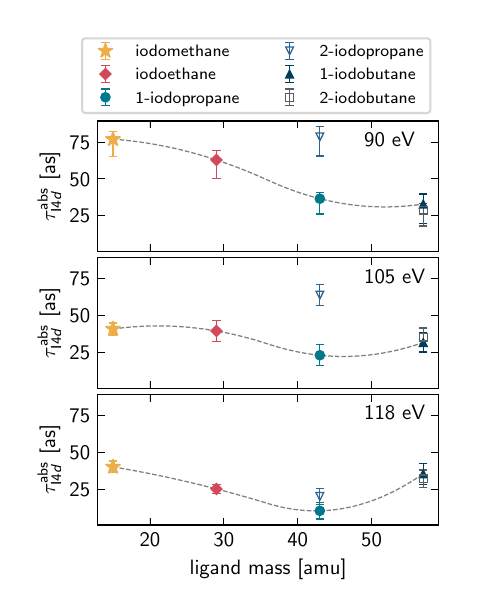}
	\caption{Correlating the absolute I$4d$ photoemission time delay with the ligand size as measured by its mass. While Ref. \cite{biswas2020probing} predicts an increase of the I$4d$ photoemission delay with ligand mass we find that $\tau^\mathrm{abs}_{\mathrm I 4d}$ tends to \textsl{decrease} (primary and secondary iodoalkanes at $90\,\mathrm{eV}$, secondary iodoalkanes at $105\,\mathrm{eV}$) or to initially decrease and go through a shallow minimum for 1-iodopropane as seen for the primary iodoalkanes at $105\,\mathrm{eV}$ and $118\,\mathrm{eV}$. The gray dashed lines have been added as a guide to the eye.}
	\label{img:mass_correlation}
\end{figure}

\citet{biswas2020probing} use their semi-classical calculations also to predict how the I$4d$ photoemission time should behave as the attached functional group is varied, and find that the total I$4d$ photoemission time nearly doubles when moving from iodoethane to 2-iodo-2,3,3-trimethylbutane ($(\mathrm{CH_3})_5\mathrm{C_2}\mathrm{I}$, see fig. 4e in \cite{biswas2020probing}), assuming an intermediate value for the fluorinated analog of iodoethane. Although the authors do not state at which excitation energy specifically this result is obtained, they predict a streaking time shift of $\sim 12\,\mathrm{as}$ for iodoethane in this calculation, which is close to their calculated data point at $105\,\mathrm{eV}$ in Fig. 3a of Ref. \cite{biswas2020probing}. Their calculated EWS time delay reported at this energy is $\sim 14\,\mathrm{as}$ (Fig. 3b of \cite{biswas2020probing}). They attribute the predicted increase from iodoethane to $(\mathrm{CH_3})_5\mathrm{C_2}\mathrm{I}$ to the difference in molecular environment the leaving photoelectron experiences, which manifests as ''an increase of the overall delay with the size of the ligand'' \cite[p. 783]{biswas2020probing}. While our study does not address $(\mathrm{CH_3})_5\mathrm{C_2}\mathrm{I}$ specifically, our results generally do not consistently show the photoemission time delay increasing as the ligand size increases. We also do not find iodoethane to have an EWS delay of $14\,\mathrm{as}$ at $105\,\mathrm{eV}$, or any other photon energy our experiment is conducted at. As fig. \ref{img:mass_correlation} shows, we find that the I$4d$ photoemission delay rather tends to \textsl{decrease} as the size of the functional group (as measured by its mass) increases. At $90\,\mathrm{eV}$ this decrease occurs consistently across iodomethane, -ethane and 1-propane, as well as between 2-iodopropane and 2-iodobutane. For the higher photon energies and the primary iodoalkanes $\tau^\mathrm{abs}_{\mathrm I 4d}$ goes through a minimum for 1-iodopropane, while for the secondary iodoalkanes we find the delay to decrease with ligand mass at $105\,\mathrm{eV}$ and increase at $118\,\mathrm{eV}$. Overall, no meaningful and consistent correlation can be established. Now, the question of whether the decrease of the I$4d$ photoemission time observed here will continue towards 2-iodo-2,3,3-trimethylbutane or whether one will
find it to rise again, in line with the prediction in \cite{biswas2020probing} can only be answered by performing the experiment on the difficult to obtain molecule. However, taking all available data (fig. \ref{img:delays}) into account, it stands to reason that the semi-classical calculation of \citet{biswas2020probing} is incorrect: their own experiment agrees with this calculation only at a single photon energy ($93\,\mathrm{eV}$) within uncertainties, and this result is replicated neither by \citet{ossiander2018absolute}, nor by us. Furthermore, the calculation by \citet{pi2018attosecond} for atomic iodine (the accuracy of which can be verified by comparison of the predicted cross-section with experimental data, e.g. \cite{nahon1991experimental, olney1998quantitative}) gives much larger values than what \citet{biswas2020probing} find for the isolated atom, corroborating our assessment. We must therefore conclude that the predicted correlation of the photoemission time delay with the size of the molecule in \cite{biswas2020probing} is simply an artifact of their calculation, and it is therefore not observed in our experiment.

\begin{figure}[ht!]
	\includegraphics{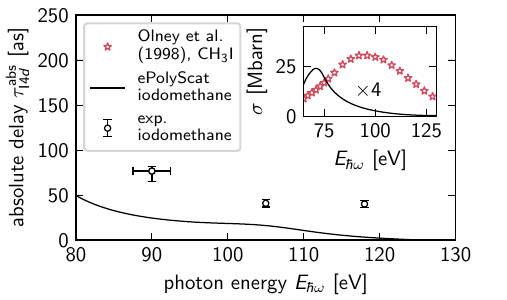}
	\caption{Comparison of our experimental result for iodomethane with a quantum scattering calculation on the HF+LDA level of theory (\texttt{ePolyScat}). The inset shows the calculated cross section in comparison to the experiment on iodomethane by \citet{olney1998quantitative}. It is easily seen that the giant resonance is not rendered adequately.}
	\label{img:lda_failure}
\end{figure}

\subsection{How sensitive is the experiment to the molecular environment?}
Another noteworthy observation is that at all studied photon energies and within their respective uncertainties 1- and 2-iodobutane have equal delays, which are very close to the atomic prediction by \citet{pi2018attosecond}. This raises the question of whether the experiment may become insensitive to the position of the Iodine atom within the molecule for larger molecules, or even insensitive to the presence and shape of the ligand overall. Furthermore, this shows that the mere \textsl{presence} of a functional group does not necessarily raise the I$4d$ photoemission delay when compared to the atomic case. Unfortunately, an accurate independent theoretical assessment of the photoemission time delay in the iodoalkanes is currently out of reach. Both \citet{ossiander2018absolute} and \citet{biswas2020probing} compare their experimental results with scattering calculations based on the Schwinger variational procedure and local density approximation (LDA) on the Hartree-Fock (HF) level of theory \cite{gianturco1994calculation,natalense1999cross, GAMESS} using the \texttt{ePolyScat} software package (see supplementary information for details), but as Fig. \ref{img:lda_failure} shows such calculations fail to produce photoionization cross-sections in agreement with experimental data even for iodomethane, discouraging comparison of calculation and our experiment. The primary reason for this failure lies in the inadequacy of the LDA as a description of the giant resonance in the I$4d\to\varepsilon f$ photoemission, as it has also previously been shown for the giant resonance in both the I$4d$ \cite{osullivan1996trends} and Xe$4d$ photoemission \cite{magrakvelidze2016attosecond}. For the larger molecules these calculations quickly become infeasible overall as they rely on a single-center partial wave expansion, which requires prohibitively large values of $\ell$ to converge as the molecule grows in size. A promising candidate for the theoretical assessment of photoemission time delays in molecules at an appropriate level of theory, however, is the $R$-matrix method with time dependence as it is developed by \citet{masin2020ukrmol}.

\section{Conclusion}
In this work we presented a systematic survey of the I$4d$ photoemission time delay in the iodoalkanes from iodomethane to 2-iodobutane. We find that the presence of different functional groups modulates the delay in a unique manner for the smaller molecules under study, while we find very similar values in the larger iodoalkanes. Overall, our results indicate that the I$4d$ photoemission time as determined via attosecond streaking spectroscopy in the gas phase does not increase with the size of the attached functional group as it has been suggested by \citet{biswas2020probing}. Generally, no clear correlation between observed delay and any measure of size of the molecule can be established from the presented data, and is evident that more experimental and especially theoretical work is needed to understand the origin and meaning of the variations observed in the time delays, for which we hope that our experimental results can serve as a useful benchmark.

\section{Author contributions}
C.A.S. conceived the study, performed the experiments and calculations, analyzed the data and wrote the first draft of the manuscript. M.P., P.F. S.J.P. and  M.F. assisted with the experiments. M.O. estimated the influence of the NVV Auger-Meitner emission on the delay retrieval, R.K. supervised the study. All authors reviewed the manuscript and contributed with valuable discussions.

\section{Acknowledgements}
The authors gratefully acknowledge valuable discussions with Andreas Westfahl and Michael Mittermair, and thank Alexander Guggenmoos, Yang Cui and Ulf Kleineberg for providing the XUV multilayer reflectors.

\bibliography{bibliography.bib}

\onecolumngrid
\pagebreak
\widetext
\begin{center}
\textbf{\large Supplementary information for: Photoemission chronoscopy of the Iodoalkanes}
\end{center}

\setcounter{equation}{0}
\setcounter{figure}{0}
\setcounter{table}{0}
\setcounter{page}{1}
\makeatletter
\renewcommand{\theequation}{S\arabic{equation}}
\renewcommand{\thefigure}{S\arabic{figure}}
\renewcommand{\bibnumfmt}[1]{[S#1]}
\renewcommand{\citenumfont}[1]{S#1}

\section{Fitting procedure}

Our delay extraction procedure is based on that of \cite{schultze2010delay} and benchmarked and described in detail in \cite{brunner2022deep}. Briefly, the streaked photoemission from a single initial state in a model atom or molecule may be approximated as
\begin{equation}
	S(E, \Delta\tau) = \left|\int \chi(t + \Delta\tau) \exp\left(-i\Phi_V(\sqrt{2E_0}, t)\right)\mathrm{e}^{-iEt}dt\right|^2,
\end{equation}
where $\Phi_V(p, t) = \int_t^\infty pA_\mathrm{NIR}(t')+\frac{1}{2}A^2_\mathrm{NIR}(t')dt'$ is the Volkov phase shift accumulated due to the action of the streaking field's vector potential $A_\mathrm{NIR}(t)$ on a photoelectron with momentum $p$ and $\chi(t)$ is the time-domain representation of the photoelectron wave packet \cite{yakovlev2010attosecond}. We parametrize $A_\mathrm{NIR}(t)$ in the time domain as
\begin{equation}
	A_\mathrm{NIR}(t; A_0, \mathcal{T}, \omega_L, \beta_L, \varphi) = A_0 \exp\left(-4 \log 2 \left(\frac{t}{\mathcal{T}}\right)^2\right)\sin\left(\omega_Lt + \frac{1}{2}\beta_Lt^2+\varphi\right),
\end{equation}
i.e. a sinusoidal carrier of frequency $\omega_L$, chirp parameter $\beta$ and ce-phase $\varphi$ and an envelope of FWHM duration $\mathcal{T}$, and $\chi(t)$ in the spectral domain ($\mathcal{F}^{-1}$ denotes the inverse Fourier transform) as
\begin{equation}
	\chi(t; \omega_0, \delta, \beta, \Delta\omega) = \mathcal{F}^{-1}\left\{e^{-4\log 2\left(\frac{\omega - \omega_0}{\Delta\omega}\right)^2}e^{i\delta(\omega - \omega_0) + \frac{1}{2}i\beta(\omega - \omega_0)^2}\right\},
\end{equation}
i.e. a Gaussian wave packet of width $\Delta\omega$ centered at $\omega_0$ with a group delay of $\delta$ and chirp parameter $\beta$. An experimental spectrogram is then modeled as
\begin{eqnarray}
	S_\mathrm{fit}(E, \Delta\tau) =
	\sum_{i}^{N_i}\sum_{j}^{N_j^{(i)}}a_{i, j}
	\left|
		\int \chi(t + \Delta\tau;\omega_\mathrm{xuv} - I_p^{i, j}, \delta_i,\beta_i, \Delta\omega)\right.\nonumber\\
		\times\left.\exp\left(-i\Phi_V(\sqrt{2(\omega_\mathrm{xuv} - I_p^{i, j})}, t; A_0, \mathcal{T}, \omega_L, \beta_L, \varphi)\right)\mathrm{e}^{-iEt}dt
	\right|^2,
	\label{rTDSE}
\end{eqnarray}
where the sum over $i$ runs over the individual streaking features in a spectrogram (in our case this is the He$1s$ and the I$4d$ emission, respectively), and the inner sum over $j$ enumerates the sticks in a stick spectrum describing the respective feature (the He$1s$ is modeled by a single stick, while the for the I$4d$ the two spin-orbit components separated by $1.7,\mathrm{eV}$ are used). The $I_p^{i, j}$ are the ionization potentials assumed for the individual sticks, $\omega_{xuv}$ is the photon energy and $a_{i, j}$ determines the strength of the streaking feature in the spectrogram. The derivative of eq. \ref{rTDSE} with respect to the XUV/NIR delay $\Delta\tau$ is fit to the experimental data using the nonlinear least squares fitting routine implemented as the \texttt{leastsq} function in the \texttt{scipy} package for \texttt{python}. Taking the derivative rejects any time-independent contributions to the spectrogram and therefore obviates the need for any kind of background subtraction. All parameters with indices as well as the photon energy are optimized in the fit. The relative photoemission delay between two streaking features is the difference of their $\delta_i$.

\begin{figure}
	\includegraphics[width = \textwidth]{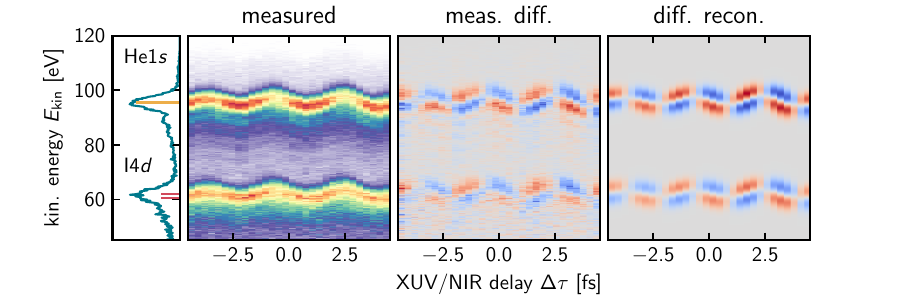}
	\caption{Example of the delay retrieval procedure, here applied to a spectrogram of a mixture of iodoethane and Helium at a photon energy of $118\,\mathrm{eV}$. The left panel shows the leftmost time slice in the spectrogram in blue and the stick spectra used to model the He$1s$ and I$4d$ feature in yellow and red, respectively. The measured spectrogram and its delay derivative are shown in the second and third panels. The rightmost panel shows the fit of eq. \ref{rTDSE} to the experimental data.}
	\label{diff_fit_example}
\end{figure}

An extensive benchmark of different delay retrieval schemes can be found in the supplementary information of \cite{brunner2022deep}. The method described here (dubbed differential rTDSE retrieval in \cite{brunner2022deep}) is very robust even in noisy conditions, where especially methods based on simply extracting the center of energy of a streaking trace fail. As demonstrated in Fig. S2 of \cite{brunner2022deep} such methods are extremely unreliable even in at good signal-to-noise conditions. The implementation of our fitting method will be published along with the experimental data.

\section{Set-up of the fitting procedure}
The He$1s$ has a binding energy of $24.6\,\mathrm{eV}$ and emits into an isolated feature in the photoelectron spectrum. The I$4d$ photoemission on the other hand is spin-orbit split into two peaks separated in energy by $\Delta E_\mathrm{so} \approx 1.7\,\mathrm{eV}$ (cf. \cite{cutler1992ligand, cutler1991relative, forbes2020photoionization}), but this splitting is not resolved in our experiment due to the large bandwidth of the XUV pulse and therefore appears as a single peak centered around a binding energy of $58.6\,\mathrm{eV}$. When applied to the experimental data, the fitting procedure is set up modeling the I$4d$ emission as a composite feature assembled from two Gaussian functions separated by $\Delta E_\mathrm{so}$, and the He$1s$ emission is modeled as a single Gaussian (see fig. \ref{diff_fit_example}). We optimize chirp, relative delay $\Delta\tau_{\mathrm{I}4d - \mathrm{He}1s}$ and energetic position of both features as well as the streaking pulse parameters, but do not permit a delay or difference in chirp between the spin-orbit components of the I$4d$ photoemission. The relative delay $\Delta\tau_{\mathrm{I}4d - \mathrm{He}1s}$ is extracted with this procedure from every spectrogram, mean and standard error are calculated per molecule and energy and used in eq. 2 of the main text to calculate the absolute $\tau^\mathrm{abs}_{\mathrm I 4d}$. At $90\,\mathrm{eV}$ we asymmetrically enlarge the reported uncertainties towards smaller delays to account for the energetic overlap of the I$4d$ primary photoemission with the NVV-Auger-Meitner electron emission occurring at kinetic energies approx. $30\,\mathrm{eV}$ and below (the $4d$ vacancy is filled by a valence electron, resulting in the emission of a valence electron).

\section{Statistical evaluation}
\begin{figure}[h]
	\includegraphics[width=\textwidth]{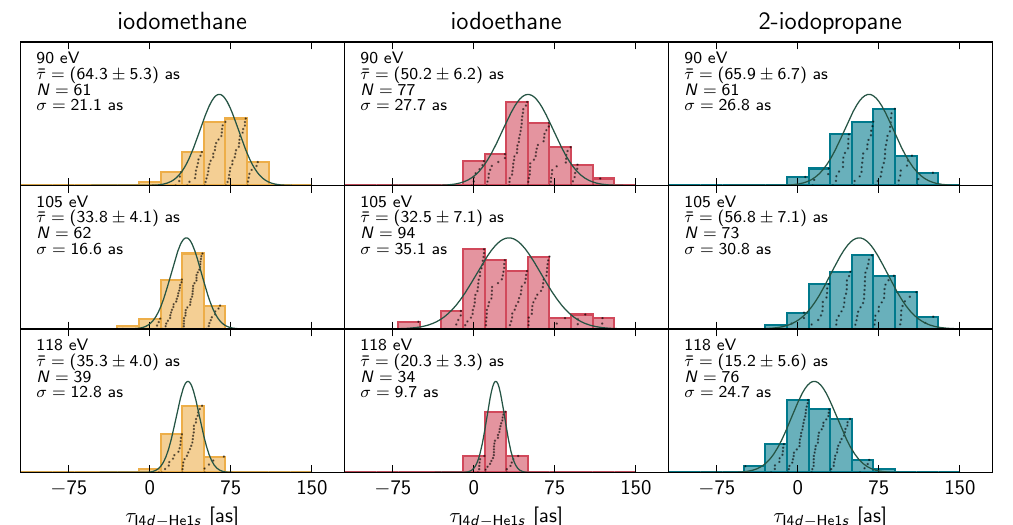}
	\caption{Statistical evaluation for iodomethane, -ethane and 2-iodopropane. Colored bars are histograms with $20\,\mathrm{as}$ bin width, black dots indicate the delays extracted from the individual measurements. Gray lines are a normal distribution with the mean and standard deviation determined from the data.}
	\label{hist_rigid}
\end{figure}

\begin{figure}[h]
	\includegraphics[width=\textwidth]{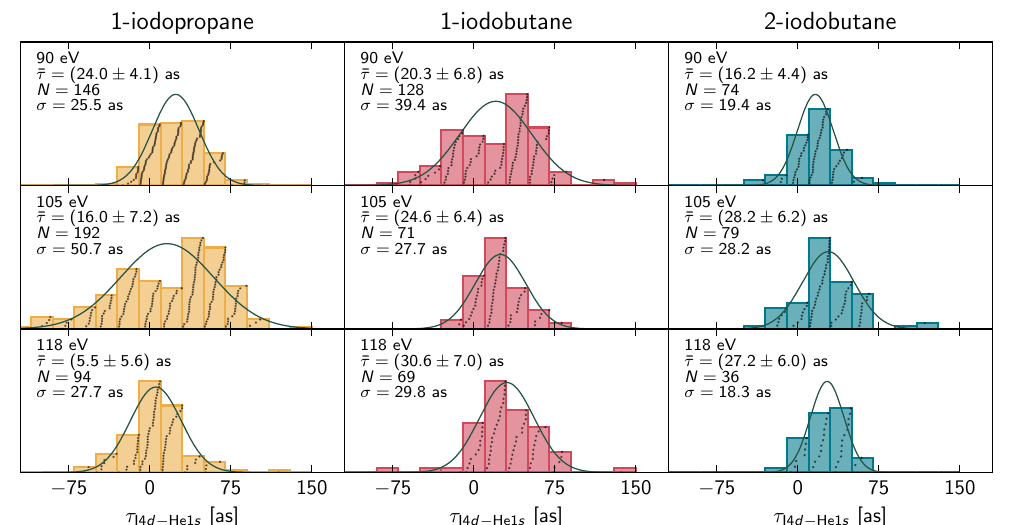}
	\caption{Statistical evaluation for 1-iodopropane, 1-iodobutane and 2-iodobutane. Colored bars are histograms with $20\,\mathrm{as}$ bin width, black dots indicate the delays extracted from the individual measurements. Gray lines are a normal distribution with the mean and standard deviation determined from the data.}
	\label{hist_floppy}
\end{figure}

For each molecule no less than 30 spectrograms are recorded over multiple days. Relative delays are extracted from the spectrograms with the fitting procedure described in the preceding section. For each molecule and energy we then determine the mean and standard deviation of the delays, reporting the $95\%$ quantile standard error $1.96\sigma / \sqrt{N}$ as the uncertainty of the mean. The entire data are compiled in figures \ref{hist_rigid} and \ref{hist_floppy}.

\subsection{Influence of the NVV Auger-Meitner emission}

\begin{figure}[h!!]
	\includegraphics[width = \textwidth]{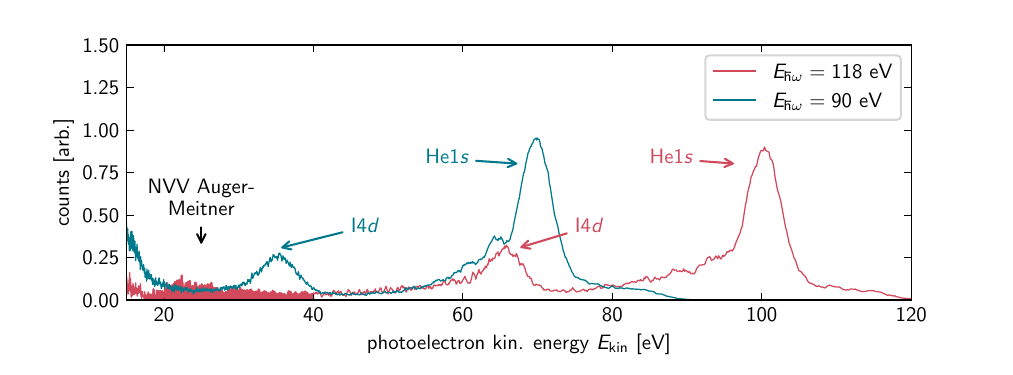}
	\caption{Streaked photoelectron spectra of 2-iodopropane at $90\,\mathrm{eV}$ (blue) and $118\,\mathrm{eV}$ (red), illustrating the overlap of the NVV Auger-Meitner emission (around $30\,\mathrm{eV}$, shaded red) with the I$4d$ emission at $90\,\mathrm{eV}$ photon energy.}
	\label{SI_static_spectra}
\end{figure}

The Auger-Meitner electron emission upon I$4d$ photoionization in Iodine compounds is found at kinetic energies around $30\,\mathrm{eV}$ \cite{karlsson1989nvv, forbes2020photoionization} and will partially overlap with the streaking trace at $90\,\mathrm{eV}$ excitation energy. Figure \ref{SI_static_spectra} illustrates the overlap.

\begin{figure}[h!!]
	\includegraphics[width = \textwidth]{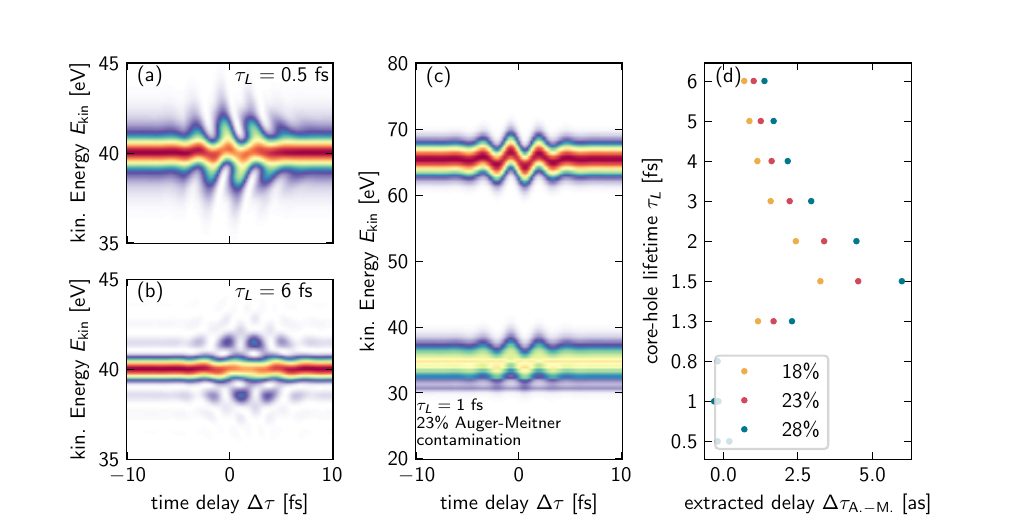}
	\caption{Estimation of the influence of the NVV Auger-Meitner emission on the time delay extraction. Panels (a) and (b) show the simulated streaked Auger-Meitner emission with different core-hole life times using the model of \cite{drescher2002time}. Panel (c) shows a complete simulated spectrogram with nominally vanishing delay with a certain configuration of Auger-Meitner contamination. Panel (d) shows how the extracted time delay varies with core-hole lifetime and amount of Auger-Meitner contamination as measured by its spectral weight relative to the modeled I$4d$ peak.}
	\label{SI_Auger_Figure}
\end{figure}

We estimate the influence of these weak features on the delay extraction by simulating the streaking of the the Auger-Meitner emission using the model from \cite{drescher2002time}, and adding this to calculated model spectrograms with nominally vanishing delays. In order to capture possible differences in the the NVV Auger-Meitner emission between molecules (especially for the larger molecules where relevant spectroscopic data is scarce) we simulate many different plausible configurations of life-times and energetic structure. Applying the fitting procedure described above we can quantify how the presence of the Auger-Meitner emission influences the extracted delays. Figure \ref{SI_Auger_Figure} shows modeled spectrograms of and including Auger-Meitner emission and how the extracted delay varies with core-hole lifetime and amount of Auger-Meitner contamination.

We report the 'worst-case' as an additional systematic uncertainty, which we find to be a shift of the delay of $6.33\,\mathrm{as}$ at a core-hole lifetime of $1.5\,\mathrm{fs}$ which is plausible from published spectroscopic data \cite{karlsson1989nvv, forbes2020photoionization}. We therefore asymmetrically extend the error bars at $90\,\mathrm{eV}$ by $-6.33\,\mathrm{as}$.

\section{XUV pulse characterization}
\begin{figure}[p]
	\includegraphics[width=\textwidth]{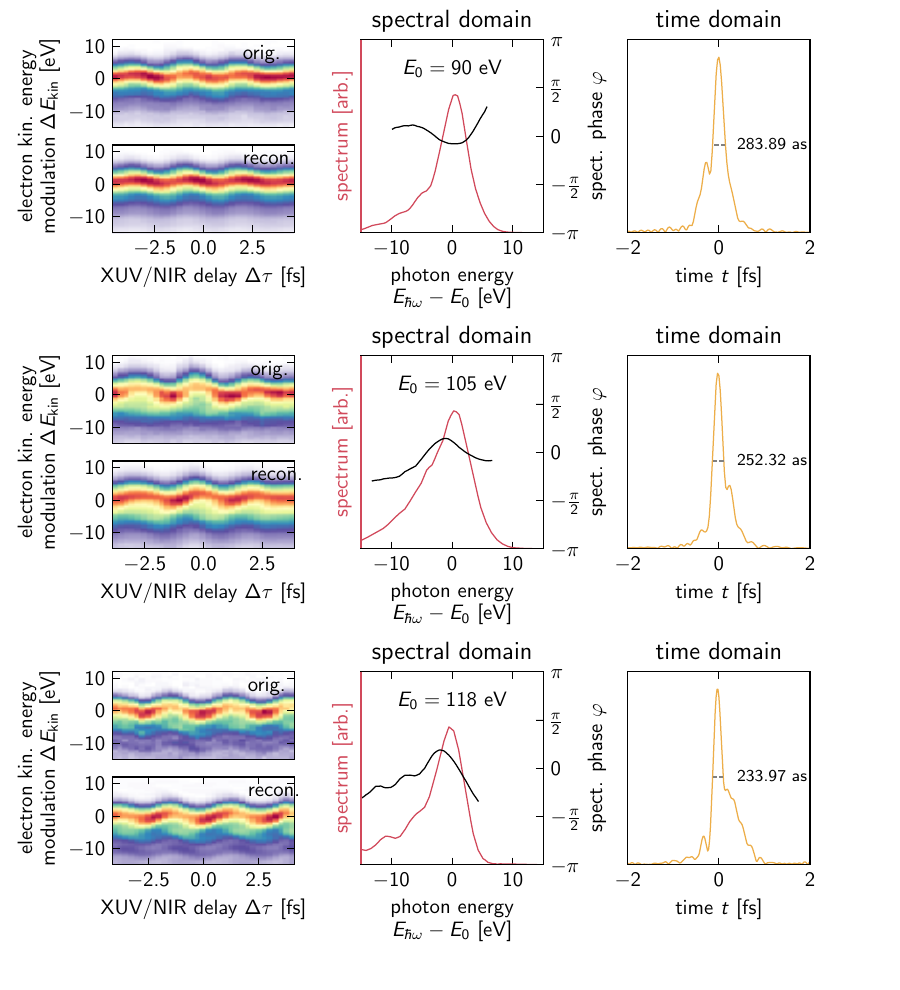}
	\caption{Attosecond pulse characterization with the extended ptychographic iterative engine (ePIE) via the He$1s$ photoemission. The exemplary spectrograms and reconstructions above show that we routinely generate isolated attosecond pulses of less than $300\,\mathrm{as}$ duration at all photon energies. The left panels show measurements and reconstructions, the center panels show the XUV pulse spectrum and spectral phase, the rightmost panels show the reconstructed pulse in the time domain.}
	\label{pulse_characterization}
\end{figure}

A convenient feature of an attosecond streaking measurement involving a noble gas is that it also characterizes the NIR streaking field as well as the attosecond XUV pulse. Figure \ref{pulse_characterization} shows exemplary spectrograms at each photon energy investigated. We reconstruct the attosecond pulse using $10^3$ iterations of the the extended ptychographic iterative engine (ePIE) \cite{lucchini2015ptychographic}. The time-domain shape of the NIR streaking field can directly be seen in the spectrograms.

\section{Scattering calculations}

We calculated the absolute (EWS) photoemission time delay for both Helium as well as iodomethane using the \texttt{ePolyScat} \cite{gianturco1994calculation,natalense1999cross} and \texttt{GAMESS} \cite{GAMESS} software packages. The input to the \texttt{ePolyScat} calculation is a Hartree-Fock calculation done with \texttt{GAMESS} \cite{GAMESS}. The \texttt{ePolyScat} calculation yields the coefficients $I^{p_i\mu_i,p_f\mu_f}_{\ell,m,\mu}(E)$ for a spherical harmonics expansion
\begin{equation}
	T^{p_i\mu_i,p_f\mu_f}_{\mu_0}(E, \hat k, R_{\hat n}) = \sum_{\ell, m,\mu} I^{p_i\mu_i,p_f\mu_f}_{\ell,m,\mu}(E) Y^{*}_{\ell,m}(\hat k) D^{1}_{\mu, -\mu_0}(R_{\hat n})
	\label{mat_ele}
\end{equation}
of the photoionization matrix elements $T^{p_i\mu_i,p_f\mu_f}_{\mu_0}(E, \hat k, R_{\hat n})$. In this expression, $\mu_0$ indicates the polarization of the impinging light ($0$ for linear, $\pm1$ for circular), $(p_i,\mu_i)$ and $(p_f,\mu_f)$ label the irreducible representations of initial- and final states, $Y^{*}_{\ell,m}(\hat k)$ are the spherical harmonics, $\hat k = (\varphi, \vartheta)$ is the direction of the photoelectron in the atomic or molecular frame and $R_{\hat n} = (\alpha, \beta, \gamma)$ is the of Euler angles that rotate the atom or molecule into the laboratory frame via the Wigner matrix $D^{1}_{\mu, -\mu_0}(R_{\hat n})$. For a transition $i \to f$ between an initial state $i$ and continuum final state $f$ driven by linearly polarized XUV light the EWS delay is
\begin{equation}
	\tau_\mathrm{EWS}(E, \hat k, R_{\hat n}) = \frac{\partial}{\partial E}\arg\left\{\sum_{\mu_i, \mu_f} T^{p_i\mu_i,p_f\mu_f}_{0}(E, \hat k, R_{\hat n}) \right\},
	\label{odd}
\end{equation}
which we average over all molecular orientations $R_{\hat n}$ while keeping $\varphi = \alpha$ and $\vartheta = \beta$ to collect only electrons in parallel with the polarization of the light.

We evaluate eqns. \ref{mat_ele} and \ref{odd} using a handful of \texttt{C++} programs and \texttt{python} scripts. Cross-sections are directly taken from the \texttt{ePolyScat} output.

\subsection{Helium}

\begin{figure}
	\includegraphics[width = \textwidth]{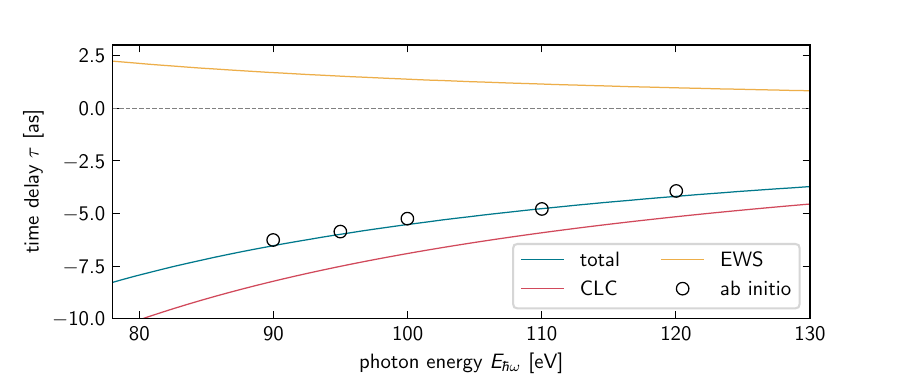}
	\caption{Calculated absolute (yellow) and total (blue) photoemission time delay for the He$1s$ photoemission, as well as the CLC contribution (red) in comparison with the \textsl{ab-initio} calculation from \cite{ossiander2017attosecond}.}
	\label{he_reference}
\end{figure}

Figure \ref{he_reference} shows the resulting EWS delay as well as the total He$1s$ photoemission delay (with the added CLC contribution \cite{pazourek2015attosecond}) and the \textsl{ab-initio} calculation from ref. \cite{ossiander2017attosecond} for comparison. The blue curve is used as the chronoscopy reference in the main text. The Sapporo-Triple-$\zeta$ basis set was used for the Hartree-Fock calculation, and the single-center expansion in \texttt{ePolyScat} was carried out up to $\ell_\mathrm{max} = 15$. The excellent agreement with the \textsl{ab-initio} calculation indicates that our procedure of determining the EWS delay from the \texttt{ePolyScat} calculations is correct.

\subsection{Iodomethane}

\begin{figure}
	\includegraphics[width = \textwidth]{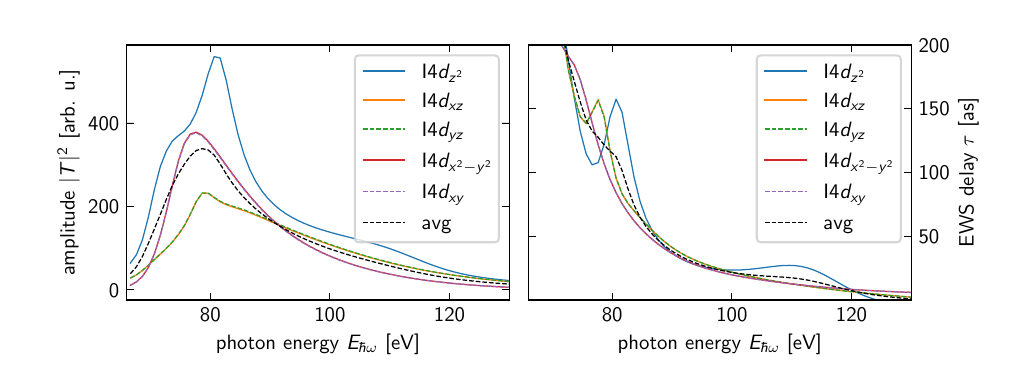}
	\caption{Orientation averaged matrix element amplitudes (left) and EWS delays (right) for the five sublevels of the I$4d$ shell in iodomethane and their respective average.}
	\label{ch3i_ePS}
\end{figure}

The Sapporo-Triple-$\zeta$ basis set was used for the geometry optimization and final Hartree-Fock calculation. We find that a $\ell_\mathrm{max}=55$ is required in \texttt{ePolyScat} to converge the single center expansion, although lower values yield practically indistinguishable delays. Figure \ref{ch3i_ePS} shows the orientational average of the matrix element amplitude (eq. \ref{mat_ele}) and EWS delays (eq. \ref{odd}) for the five sublevels of the I$4d$ in the left and right panels, respectively.

\section{Torsional barriers in the larger Iodoalkanes}

\begin{figure}
	\includegraphics{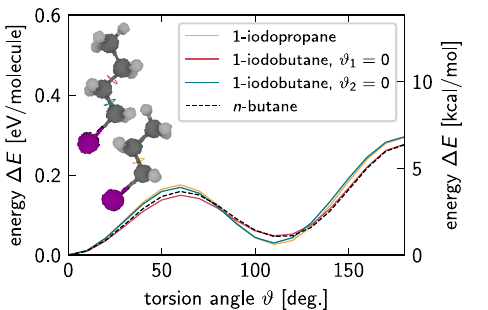}
	\caption{The rotational barriers in the larger iodoalkanes are almost identical to those in $n$-butane, permitting configurational changes at our experimental conditions. The measured I$4d$ photoemission times in these molecules therefore correspond to an average over an ensemble of molecules in different structural configurations at the time when photoemission is initiated. The torsion angles $\vartheta_i$ are enumerated starting count from the I-C bond.}
	\label{img:rotational_barriers}
\end{figure}

While a proper theoretical asessment of the I$4d$ photoemission delay is unavailable at this time, we may speculate as to why the experimentally determined $\tau^\mathrm{abs}_{\mathrm I 4d}$ converge for the largest molecules under study, as this may point towards improvements to the experiment which could yield further insight: a property that distinguishes iodomethane, -ethane and 2-iodopropane from 1-iodopropane and 1- and 2-iodobutane is the ability of the latter group to deform their molecular backbone via torsion of a C-C bond. The torsional barriers in unsubstituted alkanes are low enough to permit rapid changes of conformation at room temperature \cite{mo2011rotational}, and in the case of $n$-butane specifically an equilibrium constant of $K_\mathrm{eq} \approx 0.5$ has been determined at $T\approx 293\,\mathrm{K}$ for interconversion of its \textsl{trans} and \textsl{gauche} isomers \cite{herrebout1995enthalpy}, meaning that one can expect \textsl{trans} and \textsl{gauche} configurations to be present in roughly a 2:1 ratio at our experimental conditions. The rotational barriers in the iodoalkanes are of almost identical height and shape as in unsubstituted alkanes (see Fig. \ref{img:rotational_barriers} and supplementary information), and therefore a similar ratio of rotamer populations may be expected for the larger iodoalkanes in our experiment. Without special means to cool the gas jet in the experiment and thereby ensure that all molecules are in their lowest-energy conformation the measured $\tau^\mathrm{abs}_{\mathrm I 4d}$ must be understood as an average over an ensemble of different rotamers of the molecule under study. For the smaller molecules this may not have a large influence on the experimental result (the change in molecular structure due to rotation of a terminal methyl group in e.g. iodoethane or 2-iodopropane may not make an experimentally resolvable difference), but for the larger 1- and 2-iodobutane the differences might be significant. On the other hand, following this argumentation, rotamers of 2-iodobutane should be structurally quite similar to 2-iodopropane in the vicinity of the Iodine atom, as only torsion of the C-C bond between the second and third Carbon atom significantly changes the shape of the functional group. Yet, we measure much smaller photoemission delays in 2-iodobutane as compared to 2-iodopropane. We therefore surmise that the precise structure of the functional group probably only plays a minor role in the photoemission time accessible in our experiment, as well as that of \cite{biswas2020probing}, and find it more likely that it is another property of the molecule entirely that dominates the observed delays. The question of what exactly this property is can at this time not be answered satisfactorily though, and must be left for future work to address.

We estimate the torsional barriers in 1-iodopropane, 1-iodobutane and 2-iodobutane via a Hartree-Fock calculation in \texttt{GAMESS} \cite{GAMESS}, again using the Sapporo-Triple-$\zeta$ basis set. The molecular geometry is parametrized in the $z$-matrix representation and the dihedral angle $\vartheta_d$ of the C-C bond under study is frozen to the value corresponding to the desired conformer, then the rest of the molecular geometry is relaxed. We repeat this calculation for angles $\vartheta_d$ between $0\degree$ and $180\degree$ in 10 steps and report the difference of the final SCF energy from the lowest-energy conformer as a function of dihedral angle.

\end{document}